\documentclass[12pt]{article}
\usepackage{epsfig}
\usepackage{amssymb}
\usepackage{amsmath}
\usepackage{amsfonts}

\newcommand{\R}{\mathbb{R}}
\newcommand{\C}{\mathbb{C}}

\newcommand{\be}{\begin{equation}}
\newcommand{\ee}{\end{equation}}
\newcommand{\bea}{\begin{eqnarray}}
\newcommand{\eea}{\end{eqnarray}}

\newcommand{\kt}{\rangle}
\newcommand{\br}{\langle}

\newcommand{\ed}{\end{document}}

\newcommand{\pbr}{\prec\!}
\newcommand{\pkt}{\!\succ}

\newcommand{\bi}{\begin{itemize}}
\newcommand{\ei}{\end{itemize}}

\begin{document}

\title{Physical Meaning of Hermiticity and Shortcomings of the
Composite (Hermitian + non-Hermitian) Quantum\\ Theory of
G\"unther and Samsonov}
\author{\\
Ali Mostafazadeh
\\
\\
Department of Mathematics, Ko\c{c} University,\\
34450 Sariyer, Istanbul, Turkey\\ amostafazadeh@ku.edu.tr}
\date{ }
\maketitle

\begin{abstract}

In arXiv:0709.0483 G\"unther and Samsonov outline a
``generalization'' of quantum mechanics that involves simultaneous
consideration of Hermitian and non-Hermitian operators and
promises to be ``capable to produce effects beyond those of
standard Hermitian quantum mechanics.'' We give a simple physical
interpretation of Hermiticity and discuss in detail the
shortcomings of the above-mentioned composite quantum theory. In
particular, we show that the corresponding ``generalization of
measurement theory'' suffers from a dynamical inconsistency and
that it is by no means adequate to replace the standard
measurement theory.

\vspace{5mm}

\vspace{2mm}


\end{abstract}


In \cite{prl} we proved the following theorem.
    \begin{itemize}
    \item[] \textbf{Theorem:} {\em The lower bound on the travel
    time (upper bound on the speed) of unitary evolutions is a
    universal quantity independent of whether the evolution is
    generated by a Hermitian or non-Hermitian Hamiltonian.}
    \end{itemize}
A direct implication of this theorem, which contradicts the main
result of \cite{bender-prl-2007}, is that \emph{as far as the
Brachistochrone problem is concerned the use of non-Hermitian (in
particular ${\cal PT}$-symmetric) Hamiltonians that are capable of
generating unitary time-evolutions does not offer any advantage
over the Hermitian Hamiltonians.} This is in complete agreement
with the earlier results on the physical equivalence of the
pseudo-Hermitian (in particular ${\cal PT}$-symmetric) quantum
mechanics and the standard (Hermitian) quantum mechanics,
\cite{jpa-2003-jpb-2004b,cjp-2004}. To avoid this equivalence,
G\"unther and Samsonov \cite{gs} have recently outlined a
composite quantum theory involving both Hermitian and
quasi-Hermitian operators \cite{quasi}. They state that this
theory is a genuine generalization of the standard (Hermitian)
quantum mechanics in the sense that it allows for ``quantum
mechanical setups which are capable to produce effects beyond
those of standard Hermitian quantum mechanics'' and avoids the
consequences of the above-stated theorem of \cite{prl}. The main
purpose of the present paper is to offer a closer look into this
composite quantum theory and reveal its shortcomings. In
particular, we will show that this theory is by no means an
alternative to the standard quantum mechanics or its
pseudo-Hermitian representation.

To make our treatment precise, we first list the postulates of
pseudo-Hermitian quantum mechanics and discuss their implications.
    \begin{itemize}

    \item[(A1)] A quantum system $S$ is determined by a triplet
    $({\cal H},H,\pbr\cdot,\cdot\pkt)$, where ${\cal H}$ is a
    (separable) Hilbert space with defining inner product
    $\br\cdot|\cdot\kt$, $H:{\cal H}\to{\cal H}$ is
    (a densely defined closed) linear operator, called the
    Hamiltonian, and $\pbr\cdot,\cdot\pkt$ is a
    (positive-definite) inner product on ${\cal H}$ that may be
    different from $\br\cdot|\cdot\kt$.

    \item[(A2)] The (pure) states of $S$ are identified with the rays
    $\Lambda$ (one-dimensional subspaces) of ${\cal H}$. They may be
    represented by any nonzero element $\psi$ of
    $\Lambda$.\footnote{As the association of states $\Lambda$ to
    state vectors $\psi$ is one to infinitely many, we make a
    distinction between states and state vectors.}

    \item[(A3)] The observables  of the system are identified with
    a class of (densely defined closed) linear operators
    $O_\alpha:{\cal H}\to{\cal H}$. Upon measuring an
    observable $O_\alpha$ while the system is in a state
    $\Lambda$ one obtains a reading $\omega\in\R$ and
    the state $\Lambda$ that is prepared
    before the measurement undergoes an abrupt change into an
    eigenstate $\Lambda'$ of $O_\alpha$ with eigenvalue $\omega$.
    Alternatively, any state vector $\psi\in\Lambda$ collapses to an
    eigenvector $\psi_\omega\in\Lambda'$ of $O_\alpha$.

    \item[(A4)] All the eigenvalues (and eigenvectors) of an
    observable $O_\alpha$ are among the possible outcomes of
    a measurement of $O_\alpha$. In particular, every eigenvector of
    $O_\alpha$ may be prepared in this way.

    \item[(A5)] The outcome of a measurement is probabilistic in nature
    and the probability $P_\omega(\Lambda)$ of measuring a
    non-degenerate\footnote{The generalization to non-degenerate
    eigenvalues is straightforward. We do not consider it for
    brevity.} eigenvalue $\omega$ upon measuring
    $O_\alpha$ in a state $\Lambda$ is given by
        \be
        P_\omega(\Lambda)=\frac{|\pbr\psi,\psi_\omega\pkt|^2}{
        \sqrt{\pbr\psi,\psi\pkt\pbr\psi_\omega\psi_\omega\pkt}},
        \label{probability}
        \ee
    where $\psi\in\Lambda$ and $\psi_\omega\in\Lambda_\omega$.
    Furthermore, the expectation (mean) value of the readings
    obtained by measuring $O_\alpha$ in a state $\Lambda$ has
    the form
        \be
        \br O_\alpha\kt_\Lambda=
        \frac{\pbr\psi,O_\alpha\psi\pkt}{\pbr\psi,\psi\pkt}.
        \label{exp-val}
        \ee

    \item[(A6)] The time-evolution of an initial state $\Lambda_0$ is
    governed by the Schr\"odinger equation,
        \be
        i\hbar\frac{d}{dt}\:\psi(t)=H\psi(t),~~~~~t\geq t_0
        \label{sch-eq}
        \ee
    subject to the initial condition
        \be
        \psi(t_0)=\psi_0,
        \label{ini-cond}
        \ee
    where $\psi_0$ is any state vector belonging
    to the initial state $\Lambda_0$. The evolving state
    $\Lambda(t)$ is uniquely determined by the condition
    $\psi(t)\in\Lambda(t)$.

    \item[(A7)] The Hamiltonian $H$ is an observable.

    \end{itemize}
Note that the postulates (A1) -- (A7) do not involve the
requirement that the observables or the Hamiltonian be Hermitian
or that the time-evolution be unitary. They have however a number
of drastic consequences. Some of the most important of these are
the following.
   \begin{itemize}

    \item[(B1)] There is a dense set of state vectors that can be
    prepared for any physical measurement involving an observable
    $O_\alpha$. This follows from (\ref{exp-val}) and the
    assumption that observables are densely defined, i.e., (A3).

    \item[(B2)] The observables must
    have a real spectrum. This follows from (A4).

    \item[(B3)] The observables must
    have a complete set of eigenvectors. As discussed in
    \cite{cjp-2006}, this follows from (A3) -- (A5) and (B1).

    \item[(B4)] The observables $O_\alpha$ must be Hermitian
    with respect to the
    inner product $\pbr\cdot,\cdot\pkt$. This follows from a
    well-known theorem \cite{axler} of linear algebra saying that
    the reality
    of the expectation values of $O_\alpha$ in every state is
    equivalent to the Hermiticity of $O_\alpha$ with respect to
    $\pbr\cdot,\cdot\pkt$.

    \item[(B5)] The time evolution (operator
    $e^{-i(t-t_0)H/\hbar}$)  is unitary with respect to
    the inner product $\pbr\cdot,\cdot\pkt$. This follows from
    (A7) and (B4).

    \end{itemize}
Note that the theorem mentioned in (B4) provides a clear physical
meaning for the requirement of the Hermiticity of observables,
namely that
    \begin{center}
    \emph{\textbf{Hermiticity of an observable means reality of its
expectation values}}.\footnote{This is a more stringent condition
than the reality of the spectrum.}
    \end{center}
This seems to be overlooked by those claiming that Hermiticity is
a mathematical requirement that lacks physical meaning and hence
must be replaced by other ``physical'' conditions
\cite{bender-prl-2002-bender-ajp}.

The postulates (A1) -- (A7) of pseudo-Hermitian quantum mechanics
reduce to those of conventional (Hermitian) quantum mechanics, if
we demand that $\pbr\cdot,\cdot\pkt$ and $\br\cdot|\cdot\kt$ are
identical. The advantage of pseudo-Hermitian quantum mechanics is
that it includes the choice of $\pbr\cdot,\cdot\pkt$ as a degree
of freedom.

In pseudo-Hermitian quantum mechanics the physical Hilbert space
${\cal H}_{\rm phys}$ of a quantum system is obtained by endowing
the span of the eigenvectors of $H$ with the inner product
$\pbr\cdot,\cdot\pkt$ and Cauchy completing the resulting inner
product space into a Hilbert space. Therefore, ${\cal H}_{\rm
phys}$  is determined once one makes a choice for
$\pbr\cdot,\cdot\pkt$. The arbitrariness in the choice of
$\pbr\cdot,\cdot\pkt$ does not however lead to a genuine
generalization of Hermitian quantum mechanics. It turns out that
there is a unitary transformation $\rho:{\cal H}_{\rm
phys}\to{\cal H}$ mapping the observables $O_\alpha:{\cal H}_{\rm
phys}\to{\cal H}_{\rm phys}$ onto the Hermitian operators $o:{\cal
H}\to{\cal H}$ in such way that the probabilities of measurements
and expectation values are left invariant, \cite{cjp-2004}. This
means that a given physical system may be represented equally well
using Hermitian or pseudo-Hermitian quantum mechanics and that
there is no physical quantity capable of distinguishing between
them; they are not different theories but different
representations of the same theory.

The results of \cite{bender-prl-2007} seem to challenge the
physical equivalence of Hermitian and pseudo-Hermitian quantum
mechanics, for they imply that using a class of pseudo-Hermitian
Hamiltonians one can achieve arbitrarily fast unitary time
evolutions (which is known to be impossible in Hermitian quantum
mechanics.) The above-mentioned theorem of \cite{prl} contradicts
this claim openly \cite{prl}. In \cite{gs} the authors attempt at
devising a more general framework than the one given by (A1) --
(A7) so that they could provide an alternative interpretation for
the results of \cite{bender-prl-2007} and avoid the no-go theorem
of \cite{prl}. We will show that this attempt is unsatisfactory.

Before discussing the results of \cite{gs}, we wish to point out
that there is nothing wrong with the calculations of
\cite{bender-prl-2007}. If one defines the Hilbert space using an
inner product that renders the time evolution
unitary\footnote{This is what is done in \cite{bender-prl-2007}.},
then the calculations of \cite{bender-prl-2007} amount to
establishing the obvious fact that the travel time between
arbitrarily close initial and final states is arbitrarily small.
However, if one relaxes the requirement of the unitarity of time
evolution, one can indeed achieve arbitrarily fast evolutions
between distant states. The latter is a plausible approach in an
effective treatment of open systems.

It should also be emphasized that the possibility of non-unitary
arbitrarily fast evolutions is not surprising. A simple example of
a non-unitary evolution that achieves arbitrarily short travel
times for distant states is the following. Consider the standard
two-level system where the Hilbert space is $\C^2$ endowed with
the Euclidean inner product, and let the Hamiltonian of the system
be given by
    \be
    H:=\left(\begin{array}{cc} E & a \\ 0 & E
    \end{array}\right).
    \label{H=}
    \ee
where $E,a\in\R$ are constants. Then the state
$\Lambda_0=\Lambda(0)$ represented by the normalized state vector
$\psi_0:=\left(\begin{array}{c} 1 \\ 0\end{array}\right)$ evolves
into the state $\Lambda(t)$ represented by a normalized state
vector of the form
    \be
    \psi(t):=e^{i\varphi(t)}
    \left(\begin{array}{c} i\left(1+\frac{a^2t^2}{\hbar^2}
    \right)^{-1/2} \\ \left(1+\frac{\hbar^2}{a^2t^2}
    \right)^{-1/2}\end{array}\right),
    \label{final}
    \ee
where $\varphi(t)$ is an arbitrary real-valued function of $t$.
Now, suppose that $t$ is any positive real number and that $|a|$
is so large that $\epsilon:=\frac{\hbar}{|a|t}\ll 1$. Then
    \be
    \psi(t)=e^{i\varphi(t)}
    \left(\begin{array}{c} 0 \\ 1\end{array}\right)+
    {\cal O}(\epsilon)
    \label{lim-psi}
    \ee
In other words we can perform a spin flip in arbitrarily short
time $t>0$ provided that $|a|\gg\frac{\hbar}{t}$. The main
difference between this model and the one considered in
\cite{bender-prl-2007} is that the Hamiltonian of the latter is
Hermitian with respect to a non-Euclidean inner product. But if
one defines the Hilbert space using this new inner product, then
$S_z:=\left(\begin{array}{cc} 1 & 0\\0&-1\end{array}\right)$ is no
longer an observable and its eigenvectors $\left(\begin{array}{c}
1 \\ 0\end{array}\right)$ and $\left(\begin{array}{c} 0 \\
1\end{array}\right)$ do not represent spin states. Indeed in the
limit that the travel time (for the corresponding unitary
evolution) tends to zero, the states defined by these state
vectors coincide; there is indeed no evolution!

In \cite{gs} the authors adopt a view consisting essentially of a
superposition (simultaneous consideration) of the Hermitian and
pseudo-Hermitian representations of quantum mechanics together
with what they refer to as a ``generalization'' of the measurement
theory. The latter involves defining an observable as a conjugacy
class of a Hermitian operator. This scheme leaves many of the
physically relevant questions about theory unanswered. The most
important of these is the process of measurement itself. Suppose
that one makes a measurement of an observable in a state and reads
$\omega$. According to \cite{gs} the observable is a conjugacy
class of a Hermitian operator. But this information is not
sufficient to interpret the outcome of the measurement. Though one
might still identify $\omega$ with an eigenvalue of all the
operators in the conjugacy class (as they are isospectral), the
state obtained after the measurement cannot be an eigenstate of
all these operators. With no additional assumption on the faith of
the state after the measurement, we do not have a viable physical
theory with any sort of predictive power.

If we set up the rule, as suggested in \cite{gs}, that one must
use the Hermitian element of the conjugacy class and apply the
standard measurement theory to this operator, then one must also
deal with the issue of the uniqueness of this Hermitian element.
In principle, a conjugacy class may include more than one
Hermitian operator. A more serious problem is that if the process
of measurement is only sensitive to Hermitian operators, then what
is the reason for considering their conjugacy classes as
observables? A possible answer would be to note that the dynamics
is generated by a non-Hermitian Hamiltonian $H$. This latter
assumption leads to a host of even more severe difficulties,
particularly when one considers evolving observables, i.e., using
the Heisenberg picture. For example, the Hamiltonian $H$ will not
evolve in time, but there will be other operators in the conjugacy
class of $H$ such as the Hermitian Hamiltonian $h$ that will
change in time. Clearly, as a set of operators the conjugacy class
of $H$ will depend on time. Should now one say that the energy is
conserved? Given that one has to perform energy measurements using
$h$, the energy conservation will be lost!

Next, consider the outcome of an energy measurement. Following the
prescription of \cite{gs}, i.e., performing the measurement using
$h$, the state will collapse into an eigenstate of $h$. If one
defines the dynamics using $H$, because $h$ and $H$ do not
generally commute, repeated measurements of energy will yield
totally different values irrespective of how short the time
interval between two successive measurements is. This is simply
unacceptable, for it destroys the predictive power of the theory.
If, on the other hand, one defines the dynamics using $h$, then
one has nothing but the conventional quantum mechanics, because
the measurement is done using Hermitian operators and the dynamics
is generated by a Hermitian Hamiltonian. The presence of
non-Hermitian operators in the scheme do not affect the physical
aspects of the problem. Hence they can be safely discarded.

In summary, the so-called ``composite system'' scheme of \cite{gs}
that is offered as a way of avoiding the above-mentioned no-go
theorem of \cite{prl} suffers from very basic consistency
problems. Our attempt to resolve these problems leads to a
reduction of this composite quantum theory to the conventional
quantum mechanics (equivalently to its pseudo-Hermitian
representation). This is because a consistent treatment of a
physical system that undergoes both time-evolution and measurement
requires the use of a Hamiltonian and an observable that are both
Hermitian with respect to the same positive-definite inner
product. Using observables that are Hermitian together with a
Hamiltonian that is non-Hermitian leads to a dynamical
inconsistency reminiscent of the inconsistency
\cite{cjp-2004,comment} arising from the identification of
observables in ${\cal PT}$-symmetric quantum mechanics with the
${\cal CPT}$-symmetric operators with real spectrum
\cite{bender-prl-2002-bender-ajp}.

\ed

\ed